\documentclass[aps,pra,floatfix,nofootinbib,twocolumn]{revtex4}
\usepackage{amsmath}
\usepackage{amssymb}
\usepackage{bm}
\usepackage{epsfig}

\begin{document}

\title{Vector theory of gravity in Minkowski space-time: flat Universe
without black holes}
\author{Anatoly A. Svidzinsky}
\affiliation{Department of Physics, Institute for Quantum Studies, Texas A\&M University,
College Station TX 77843-4242 \\ asvid@jewel.tamu.edu}
\date{\today }

\begin{abstract}
We propose a new classical theory of gravity which is based on the principle
of equivalence and assumption that gravity, similarly to electrodynamics, is
described by a vector field in Minkowski space-time. We show that such
assumptions yield a unique theory of gravity; it passes all available tests
and free of singularities such as black holes. In the present theory,
gravity is described by four equations which have, e.g., exact analytical
solution for arbitrary static field. For cosmology our equations give
essentially the same evolution of the Universe as general relativity.
Predictions of our theory can be tested within next few years making more
accurate measurement of the time delay of radar signal traveling near the
Sun or by resolving the supermassive object at the center of our Galaxy with
VLBA. If general relativity is correct we must see a steady shadow from a
black hole at the Galactic center. If the present theory is right then
likely the shadow will appear and disappear periodically with a period of
about $20$ min as we predicted in JCAP 10 (2007) 018. Observation of such
oscillations will also provide evidence for dark matter axion with mass in
meV range.
\end{abstract}

\maketitle

\section{Introduction}

In 1915 Albert Einstein \cite{Eins15} completed the general theory of
relativity which then became an accepted theory of gravity. In general
relativity the space-time geometry $g_{ik}$ (metric tensor) is the
gravitational field described by the action%
\begin{equation}
I_{\text{GR}}=-\frac{c^{3}}{16\pi G}\int d^{4}x\sqrt{-g}g^{ik}R_{ik}-\int
\rho \sqrt{g_{ik}\frac{dx^{i}}{dt}\frac{dx^{k}}{dt}}d^{4}x,  \label{aGR}
\end{equation}%
where $G$ is the gravitational constant and $c$ is the speed of light. The
second term in Eq. (\ref{aGR}) describes interaction between gravitational
field and matter with the rest mass density $\rho (t,\mathbf{r})$. Variation
of (\ref{aGR}) with respect to $g_{ik}$ yields Einstein equations%
\begin{equation}
R_{ik}=\frac{8\pi G}{c^{4}}\left( T_{ik}-\frac{1}{2}g_{ik}T\right) ,
\label{i1}
\end{equation}%
where $R_{ik}$ is the Ricci tensor and $T_{ik}$ is the energy-momentum
tensor of matter.

Einstein equations (\ref{i1}) are a consequence of the postulate that
space-time geometry $g_{ik}$ is gravitational field, and thus we live in
curved space-time. From our point of view this postulate is counterintuitive
because the well-tested Standard Model of particle physics (which includes
the electroweak theory and quantum chromodynamics) is a field theory in
Minkowski space-time. We believe it is unlikely that the fourth fundamental
interaction, gravity, should be described by a theory which such
dramatically different from the Standard model as general relativity. One
should mention that so far general relativity was tested only at weak
gravitational field \cite{Will06} and thus it is not a theory fully
confirmed experimentally.

General relativity also predicts existence of singularities such as black
holes when a massive star collapses into a point with zero volume and
infinite matter density. One can argue that general relativity becomes
invalid in the vicinity of singularities and a quantum theory of gravity
will remove them. In contrast, the present theory is free of such
singularities at the classical level.

Motivated by the well-tested Standard model, here we propose a new classical
theory of gravity which is a field theory in Minkowski space-time and based
on the principle of equivalence. We postulate that space-time we live is
flat Minkowski space-time and matter does not affect space-time geometry.
Also we assume, similarly to electrodynamics, that gravitational field is a $%
4-$vector $A_{k}$ which lives in Minkowski space-time.

Next we derive action for $A_{k}$. The principle of equivalence states that
motion of test particles in the fixed gravitational field $A_{k}$ in
Minkowski space-time is equivalent to motion in curved space-time with a
metric $f_{ik}$ which is determined by the field $A_{k}$. Thus, the
equivalent metric $f_{ik}$ must be a functional of the vector field $A_{k}$.
In Appendix A we show that the principle of equivalence gives the following
unique answer for $f_{ik}$%
\begin{equation}
f_{ik}=\eta _{ik}e^{A^{2}}-\frac{2A_{i}A_{k}}{A^{2}}\sinh (A^{2}),
\label{fa1}
\end{equation}%
where $\eta _{ik}=$diag$(1,-1,-1,-1)$ is Minkowski metric tensor and 
\begin{equation}
A^{2}=\eta ^{ik}A_{i}A_{k}.
\end{equation}

General covariance (invariance under general coordinate transformations) is
a mathematical device used to implement the principle of equivalence \cite%
{Wein72}. Thus, action for gravitational field written in terms of the
equivalent metric $f_{ik}$ must have general covariant form given by $I_{%
\text{GR}}$, where $I_{\text{GR}}$ is the action of general relativity (\ref%
{aGR}) in which $g_{ik}$ is replaced by $f_{ik}$. As a result, we obtain the
following expression for the action%
\begin{equation}
I=-\frac{c^{3}}{16\pi G}\int d^{4}x\sqrt{-f}\tilde{f}^{ik}R_{ik}-\int \rho 
\sqrt{f_{ik}\frac{dx^{i}}{dt}\frac{dx^{k}}{dt}}d^{4}x,  \label{fa2}
\end{equation}%
where $f=\det (f_{ik})=-e^{2A^{2}}$ and $\tilde{f}^{ik}$ is the tensor
inverse to $f_{ik}$ ($f_{im}\tilde{f}^{km}=\delta _{i}^{k}$). The first term
in Eq. (\ref{fa2}) is Einstein-Hilbert action in which the Ricci tensor $%
R_{ik}$ is formed from the field $f_{ik}$.

Action (\ref{fa2}) is a functional of the gravitational field $A_{k}$.
Variation of (\ref{fa2}) with respect to $A_{k}$ yields four equations for
gravitational field. In general relativity all components of $f_{ik}$ are
treated as independent under variation of the action (\ref{fa2}). In the
present theory this is not the case due to constraint imposed by Eq. (\ref%
{fa1}).

Action (\ref{fa2}) is written in Minkowski metric in covariant form, has no
free parameters and serves as a foundation of the present theory of gravity.
Our derivation of the action (\ref{fa2}) is unique and, hence, the vector
theory of gravity which obeys the principle of equivalence is also unique.%
\footnote{%
One should mention that, similarly to the vector potential in
electrodynamics, $A_{k}$ is not observable directly. Hence, $A_{k}$ can be
any vector field which gives physically reasonable equivalent metric $f_{ik}$%
. Because $f_{ik}$ depends on $A_{k}$ quadratically the equivalent metric (%
\ref{fa1}) remains real also for pure imaginary $A_{k}$. If $A_{k}$ is real
(pure imaginary) then always $f_{00}\leq 1$ ($f_{00}>1$). We will allow both
real and pure imaginary $A_{k}$. If gravitational field realizable in nature
corresponds only to real $A_{k}$, then solutions of our equations for
particular physical problems will automatically yield real $A_{k}$. This is,
e.g., the case for static gravitational field produced by positive masses
(see Eqs. (\ref{x4}) and (\ref{x5}) which yield $f_{00}\leq 1$).}

In Section \ref{test} we show that current theory passes all available
tests. At strong field our theory substantially deviates from general
relativity and yields no black holes. For cosmology the present theory gives
essentially the same evolution of the Universe as general relativity.

A remarkable feature of our theory is that equations for gravitational field
can be solved analytically for arbitrary static mass distribution (see Sec. %
\ref{static}). If point masses are located at $\mathbf{r}_{1}$, $\mathbf{r}%
_{2}$, \ldots\ $\mathbf{r}_{N}$ then exact solution is 
\begin{equation}
f_{ik}=\left( 
\begin{array}{cccc}
e^{2\phi } & 0 & 0 & 0 \\ 
0 & -e^{-2\phi } & 0 & 0 \\ 
0 & 0 & -e^{-2\phi } & 0 \\ 
0 & 0 & 0 & -e^{-2\phi }%
\end{array}%
\right) ,  \label{x4}
\end{equation}%
where 
\begin{equation}
\phi (\mathbf{r})=-\frac{m_{1}}{|\mathbf{r}-\mathbf{r}_{1}|}-\ldots -\frac{%
m_{N}}{|\mathbf{r}-\mathbf{r}_{N}|}  \label{x5}
\end{equation}%
and $m_{k}$ ($k=1,\ldots ,N$) are constants determined by the value of
masses.

Solution (\ref{x4}) is free of black holes. For a star of mass $M$ and
radius $R$ Eq. (\ref{x5}) reduces to $\phi (r)=-GM/c^{2}r$ ($r\geq R$) and
using Eq. (\ref{u12}) for energy conservation we obtain that escape velocity
for a particle from the stellar surface is 
\begin{equation}
v=c_{s}\sqrt{1-e^{2\phi (R)}},  \label{x6}
\end{equation}%
where $c_{s}=ce^{2\phi (R)}$ is the speed of light at the stellar surface
(see Eq. (\ref{u21})). Eq. (\ref{x6}) shows that escape velocity is always
smaller then $c_{s}$ ($c_{s}\leq c$). In addition, solution (\ref{x4})
predicts that starts do not collapse into a point singularity but rather
form stable compact objects with no event horizon and finite gravitational
redshift \cite{Robe99}.

In recent years, the evidence for the existence of ultra-compact
supermassive objects at centers of galaxies has become very strong. It is
important to note that present solution (\ref{x4}) not only argues that such
objects are not black holes, but also can explain quantitatively their
observed properties (see Sec. \ref{dark} and Ref. \cite{Svid07}).

\section{Equations for gravitational field in Minkowski space-time}

Here we obtain equations for gravitational field from the action (\ref{fa2}%
). In the rest part of the paper raising and lowering of indexes is carried
out using Minkowski tensor $\eta ^{ik}=$diag$(1,-1,-1,-1)$. It is convenient
to introduce new independent functions, a scalar 
\begin{equation}
\phi =-\frac{A^{2}}{2}  \label{x1}
\end{equation}%
and replace $A_{k}$ by a unit vector which we also will denote as $A_{k}$ 
\begin{equation}
A_{k}\rightarrow \frac{A_{k}}{\sqrt{\eta ^{lm}A_{l}A_{m}}}.  \label{x2}
\end{equation}%
The new vector $A_{k}$ obeys the normalization constraint%
\begin{equation}
A_{k}A^{k}=1.
\end{equation}%
Because expression under the square root in Eq. (\ref{x2}) can be both
positive or negative the unit vector $A_{k}$ is a real or pure imaginary
vector. Observable quantities, of course, are always real.

We treat spatial components $A_{\alpha }$ ($\alpha =1,2,3$) as independent,
while 
\begin{equation}
A_{0}=\sqrt{1+A_{1}^{2}+A_{2}^{2}+A_{3}^{2}}.
\end{equation}%
In terms of new independent functions, $\phi $ and the unit vector $A_{k}$,
the equivalent metric (\ref{fa1}) has the form%
\begin{equation}
f_{ik}=\eta _{ik}e^{-2\phi }+2A_{i}A_{k}\sinh (2\phi ),\quad \sqrt{-f}%
=e^{-2\phi }  \label{x3a}
\end{equation}%
and the inverse tensor is%
\begin{equation}
\tilde{f}^{ik}=\eta ^{ik}e^{2\phi }-2A^{i}A^{k}\sinh (2\phi ).  \label{x3}
\end{equation}%
One can find variation of the action (\ref{fa2}) using formulas 
\begin{equation*}
\delta \int d^{4}x\sqrt{-f}\tilde{f}^{ik}R_{ik}=\int d^{4}x\sqrt{-f}\left(
R_{ik}-\frac{1}{2}f_{ik}R\right) \delta \tilde{f}^{ik},
\end{equation*}%
\begin{equation*}
\delta f_{ik}=-f_{im}f_{kl}\delta \tilde{f}^{ml}
\end{equation*}%
and expressing variation of $\delta \tilde{f}^{ik}$ in terms of $\delta \phi 
$ and $\delta A^{\alpha }$. Variation of (\ref{fa2}) with respect to $\phi $
and $A^{\alpha }$ yields the following four equations in Minkowski space-time%
\begin{equation}
R_{ik}A^{k}=\frac{8\pi G}{c^{4}}\left( T_{ik}-\frac{1}{2}f_{ik}T\right)
A^{k},  \label{e2}
\end{equation}%
where the Ricci tensor $R_{ik}$ and Christoffel symbols $\Gamma _{ik}^{l}$
are formed from the equivalent metric $f_{ik}$%
\begin{equation}
R_{ik}=\frac{\partial \Gamma _{ik}^{l}}{\partial x^{l}}-\frac{\partial
\Gamma _{il}^{l}}{\partial x^{k}}+\Gamma _{ik}^{m}\Gamma _{ml}^{l}-\Gamma
_{il}^{m}\Gamma _{km}^{l},  \label{Rt}
\end{equation}%
\begin{equation}
\Gamma _{ik}^{l}=\frac{1}{2}\tilde{f}^{lm}\left( \frac{\partial f_{mi}}{%
\partial x^{k}}+\frac{\partial f_{mk}}{\partial x^{i}}-\frac{\partial f_{ik}%
}{\partial x^{m}}\right) ,  \label{Cs}
\end{equation}%
$T_{ik}$ is the energy-momentum tensor of matter%
\begin{equation}
T_{ik}=\frac{\rho c^{2}}{\gamma }e^{2\phi }f_{ia}f_{kb}u^{a}u^{b},\quad T=%
\tilde{f}^{ik}T_{ik}=\frac{\rho c^{2}}{\gamma }e^{2\phi },  \label{s3}
\end{equation}%
$\rho $ is the rest mass density, $u^{i}$ is the particle $4-$velocity 
\begin{equation}
u^{i}=\frac{\gamma }{c}\frac{dx^{i}}{dt}\equiv \gamma \left( 1,\frac{%
V^{\alpha }}{c}\right) ,  \label{e3}
\end{equation}%
$V^{\alpha }=dx^{\alpha }/dt$ ($\alpha =1,2,3$) is three dimensional
velocity of particle ($V^{2}=V_{1}^{2}+V_{2}^{2}+V_{3}^{2}$), and%
\begin{equation}
\gamma =\frac{c}{\sqrt{\left( c^{2}-V^{2}\right) e^{-2\phi }+2\left(
A_{0}c+A_{\alpha }V^{\alpha }\right) ^{2}\sinh (2\phi )}}  \label{e4}
\end{equation}%
is the $\gamma -$factor in the presence of gravitational field.

Covariant Eqs. (\ref{e2}) for the scalar $\phi $ and the unit vector $A_{k}$
are the main equations of the present theory of gravity. They are written in
Minkowski metric which means that raising and lowering of indexes is carried
out using Minkowski tensor. In our theory motion of particles in
gravitational field is described by Eq. (\ref{r10}) identical to those in
general relativity in metric $f_{ik}$. We obtain equations of particle
motion in Appendix \ref{motion}.

One should note that because $\phi $ and $A_{k}$ are not observable directly
they may not be smooth functions and components of $A_{k}$ can have jumps at
points where $\phi =0$. However, equivalent metric $f_{ik}$ constructed from 
$\phi $ and $A_{k}$ must be continuous and smooth.

Present Eqs. (\ref{e2}) have exact analytical solutions in the class of
functions for which only one component of $A_{k}$ (e.g. component $m$) is
nonzero and $\phi $ is independent of $x^{m}$. In particular, solution $%
A_{k}=[1,0,0,0]$, $\phi =\phi (\mathbf{r})$ describes static field, while
solution with $A_{\alpha }=i$ corresponds to a gravitational wave with
polarization along the $\alpha -$axis. We discuss such solutions next.

\section{Gravitational waves of arbitrary amplitude}

In general relativity there exist transverse gravitational waves of two
polarizations \cite{Land95}. Here we show that the same situation takes
place in the present theory of gravity. Let us consider a plane
gravitational wave in free space. Then gravitational field obeys equations 
\begin{equation}
R_{ik}A^{k}=0.  \label{v1}
\end{equation}%
Exact solution of Eqs. (\ref{v1}) for arbitrary strength of gravitational
field is given by 
\begin{equation}
A_{k}=[0,0,i,0],\quad \phi =\phi (t,x,z)  \label{v2}
\end{equation}%
for a wave of $y$ polarization propagating in the $x-z$ plane and%
\begin{equation}
A_{k}=[0,0,0,i],\quad \phi =\phi (t,x,y)  \label{v3}
\end{equation}%
for a wave of $z$ polarization propagating in the $x-y$ plane. For both
polarizations, $\phi $ is a function satisfying the linear wave equation 
\begin{equation}
\square \phi =0\text{,}\quad \text{where}\quad \square \equiv \Delta -\frac{1%
}{c^{2}}\frac{\partial ^{2}}{\partial t^{2}},  \label{v4}
\end{equation}%
that is wave propagates with the speed of light $c$.

For example, for $y$ polarization (\ref{v2}) the equivalent metric (\ref{x3a}%
) has the form%
\begin{equation}
f_{ik}=\left( 
\begin{array}{cccc}
e^{-2\phi } & 0 & 0 & 0 \\ 
0 & -e^{-2\phi } & 0 & 0 \\ 
0 & 0 & -e^{2\phi } & 0 \\ 
0 & 0 & 0 & -e^{-2\phi }%
\end{array}%
\right)  \label{v5}
\end{equation}%
and the corresponding Ricci tensor is%
\begin{equation}
R_{ik}=\square \phi \left( 
\begin{array}{cccc}
-1 & 0 & 0 & 0 \\ 
0 & 1 & 0 & 0 \\ 
0 & 0 & -e^{4\phi } & 0 \\ 
0 & 0 & 0 & 1%
\end{array}%
\right) -2\frac{\partial \phi }{\partial x^{i}}\frac{\partial \phi }{%
\partial x^{k}}.  \label{v6}
\end{equation}%
Obviously Eqs. (\ref{v1}) are satisfied provided $\phi (t,x,z)$ obeys the
wave equation (\ref{v4}). Combining $y$ and $z$ polarizations one can find a
general solution for arbitrary plane wave propagating along the $x-$axis: 
\begin{equation}
A_{k}=i[0,0,\sin \alpha ,\cos \alpha ],\quad \phi =\phi (t,x),\quad \alpha
=\alpha (t,x).  \label{v7}
\end{equation}%
The corresponding equivalent metric is given by%
\begin{equation}
f_{ik}=e^{-2\phi }\eta _{ik}-\sinh (2\phi )\left( 
\begin{array}{cccc}
0 & 0 & 0 & 0 \\ 
0 & 0 & 0 & 0 \\ 
0 & 0 & 2\sin ^{2}\alpha & \sin (2\alpha ) \\ 
0 & 0 & \sin (2\alpha ) & 2\cos ^{2}\alpha%
\end{array}%
\right) .  \label{m1}
\end{equation}%
For the gravitational field (\ref{m1}) Eqs. (\ref{v1}) lead to the following
equations for $\phi $ and $\alpha $%
\begin{equation}
\square \phi +\sinh (4\phi )\left[ \frac{1}{c^{2}}\left( \frac{\partial
\alpha }{\partial t}\right) ^{2}-\left( \frac{\partial \alpha }{\partial x}%
\right) ^{2}\right] =0,  \label{m2}
\end{equation}%
\begin{equation}
\sinh (2\phi )\square \alpha +4\cosh (2\phi )\left( \frac{\partial \alpha }{%
\partial x}\frac{\partial \phi }{\partial x}-\frac{1}{c^{2}}\frac{\partial
\alpha }{\partial t}\frac{\partial \phi }{\partial t}\right) =0  \label{m3}
\end{equation}%
which have exact analytical solution in the form 
\begin{equation}
\phi =\phi (x-ct),\quad \alpha =\alpha (x-ct).  \label{v8}
\end{equation}%
Eqs. (\ref{m1}) and (\ref{v8}) give a general solution for gravitational
wave of arbitrary amplitude propagating along the positive $x-$direction.

One should note that in the weak field limit solutions obtained above reduce
to those of general relativity. Indeed, in the weak field limit one can omit
nonlinear terms in Eq. (\ref{v6}). The linearized $R_{ik}$ satisfies
Einstein equations in free space $R_{ik}=0$ and, thus, the present solutions
are also solutions in general relativity. However for strong field this is
not the case.

In Appendix C we show that in our theory radiation of weak gravitational
waves is given by the same formula as in general relativity. However, the
present vector theory of gravity is not equivalent to general relativity in
the weak field limit and there are situations when two theories give
different answers even for weak field. As an example, let us consider
solution of Eqs. (\ref{m2}) and (\ref{m3}) in the form%
\begin{equation}
\alpha =\omega t,\quad \phi =\phi (x),  \label{m4}
\end{equation}%
where $\omega $ is a constant. Then Eqs. (\ref{m2}) and (\ref{m3}) yield the
following equation for $\phi (x)$

\begin{equation}
\frac{\partial ^{2}\phi }{\partial x^{2}}+k^{2}\sinh (4\phi )=0,\quad k=%
\frac{\omega }{c}  \label{m5}
\end{equation}%
which has exact analytical solution in terms of the Jacobi elliptic function
sn$(u,m)$%
\begin{equation}
\sinh (2\phi )=C\text{sn}\left( 2kx+\theta ,-C^{2}\right) ,  \label{m6}
\end{equation}%
where $C$ and $\theta $ are arbitrary constants.

In the weak field limit ($|\phi |\ll 1$) Eq. (\ref{m5}) reduces to a simple
harmonic oscillator equation 
\begin{equation}
\frac{\partial ^{2}\phi }{\partial x^{2}}+4k^{2}\phi =0  \label{m7}
\end{equation}%
and has a solution 
\begin{equation}
2\phi =C\sin (2kx+\theta ).  \label{m8}
\end{equation}%
Solution given by Eqs. (\ref{m4}), (\ref{m6}) and (\ref{m8}) describes a
standing helical gravitational wave for which $f_{ik}=\eta _{ik}$ at nodal
points and the unit vector 
\begin{equation}
A_{k}=i[0,0,\sin (\omega t),\cos (\omega t)]
\end{equation}%
rotates in the $y-z$ plane with angular frequency $\omega $.

Even in the weak field limit solution (\ref{m8}) is not a solution of
Einstein equations%
\begin{equation}
R_{ik}=0.  \label{m9}
\end{equation}%
Indeed, for metric (\ref{m1}) with $\alpha =ckt$ and $2\phi =C\sin
(2kx+\theta )$ the Ricci tensor in the first order in $\phi $ reads%
\begin{equation}
R_{ik}=4k^{2}\phi \left( 
\begin{array}{cccc}
1 & 0 & 0 & 0 \\ 
0 & -1 & 0 & 0 \\ 
0 & 0 & 0 & 0 \\ 
0 & 0 & 0 & 0%
\end{array}%
\right) .
\end{equation}%
Because $R_{00}$ and $R_{11}$ do not vanish our solution (\ref{m8}) disobeys
Einstein equations (\ref{m9}) even for weak field. However, present
equations (\ref{v1})%
\begin{equation*}
R_{22}\sin \alpha +R_{23}\cos \alpha =0,
\end{equation*}%
\begin{equation*}
R_{33}\cos \alpha +R_{23}\sin \alpha =0
\end{equation*}%
are automatically satisfied because they do not involve $R_{00}$ and $R_{11}$%
.

Such an example demonstrates that our theory of gravity is not equivalent to
general relativity in the weak field limit (see also Appendix D). This is
expected because a vector theory can not be equivalent to a tensor theory.
However, in the weak field regimes for which general relativity was tested
(such as radiation of weak gravitational waves and weak static field) both
theories give the same answer.

\section{Static gravitational field}

\label{static}

Here we consider gravitational field produced by rest matter with density $%
\rho (\mathbf{r})$. In this case the energy-momentum tensor (\ref{s3}) has
only one nonzero component $T_{00}=\rho c^{2}e^{5\phi }$ and Eqs. (\ref{e2})
have the following exact analytical solution%
\begin{equation}
A_{k}=[1,0,0,0],  \label{s11}
\end{equation}%
and $\phi (\mathbf{r})$ obeys the equation%
\begin{equation}
\Delta \phi =\frac{4\pi G}{c^{2}}e^{\phi }\rho .  \label{u13}
\end{equation}%
For solution (\ref{s11}) the equivalent metric is given by 
\begin{equation}
f_{ik}=\left( 
\begin{array}{cccc}
e^{2\phi (\mathbf{r})} & 0 & 0 & 0 \\ 
0 & -e^{-2\phi (\mathbf{r})} & 0 & 0 \\ 
0 & 0 & -e^{-2\phi (\mathbf{r})} & 0 \\ 
0 & 0 & 0 & -e^{-2\phi (\mathbf{r})}%
\end{array}%
\right)  \label{s10}
\end{equation}%
and the corresponding Ricci tensor is 
\begin{equation}
R_{ik}=\Delta \phi \left( 
\begin{array}{cccc}
e^{4\phi } & 0 & 0 & 0 \\ 
0 & 1 & 0 & 0 \\ 
0 & 0 & 1 & 0 \\ 
0 & 0 & 0 & 1%
\end{array}%
\right) -2\frac{\partial \phi }{\partial x^{i}}\frac{\partial \phi }{%
\partial x^{k}}.
\end{equation}%
Because $R_{\alpha 0}=0$ Eqs. (\ref{e2}) are automatically satisfied for $%
i=1,2,3$.

In Newtonian limit Eq. (\ref{u13}) reduces to $\Delta \phi =4\pi G\rho
/c^{2} $ and, thus, $c^{2}\phi (\mathbf{r})$ has a meaning of gravitational
potential.

Solution (\ref{s10}) is free of black holes for any mass distribution and
field strength. For a point mass $M$ located at $r=0$ Eq. (\ref{u13}) leads
to $c^{2}\Delta \phi =4\pi GM\delta (\mathbf{r})$ and has a solution $\phi
=-GM/c^{2}r$. For $N$ point masses at $\mathbf{r}_{1},$ \ldots $,$ $\mathbf{r%
}_{N}$ Eq. (\ref{u13}) yields%
\begin{equation}
\Delta \phi =4\pi \left[ m_{1}\delta (\mathbf{r}_{1})+\ldots +m_{N}\delta (%
\mathbf{r}_{N})\right] ,  \label{r6}
\end{equation}%
where $m_{1}$, \ldots , $m_{N}$ are positive constants. Solution of Eq. (\ref%
{r6}) is 
\begin{equation}
\phi (\mathbf{r})=-\frac{m_{1}}{|\mathbf{r}-\mathbf{r}_{1}|}-\ldots -\frac{%
m_{N}}{|\mathbf{r}-\mathbf{r}_{N}|}.  \label{r7}
\end{equation}

Next we consider motion of a particle with rest mass $m$ in static
gravitational field $\phi (\mathbf{r})$. Equation of particle motion in
general case is obtained in Appendix B. Eq. (\ref{r10}) for field (\ref{s10}%
) reduces to%
\begin{equation}
\frac{d(e^{2\phi }\gamma )}{dt}=0,  \label{u12}
\end{equation}%
\begin{equation}
\frac{d\left( \gamma e^{-2\phi }\mathbf{V}\right) }{dt}=-\gamma c^{2}\left[
e^{2\phi }+\frac{V^{2}}{c^{2}}e^{-2\phi }\right] \nabla \phi ,  \label{u11}
\end{equation}%
where $\nabla \phi =\partial \phi /\partial \mathbf{r}$, $\mathbf{r}%
=x^{\alpha }$, $\mathbf{V}=\partial \mathbf{r}/\partial t$ is the particle
velocity and 
\begin{equation}
\gamma =\frac{e^{-\phi }}{\sqrt{1-\frac{V^{2}}{c^{2}}e^{-4\phi }}}.
\label{u14}
\end{equation}%
One can also find equation of particle motion (\ref{u11}) directly from
Lagrange's equation $\frac{d}{dt}\frac{\partial L}{\partial \mathbf{V}}=%
\frac{\partial L}{\partial \mathbf{r}},$ where the Lagrangian (\ref{r12})
for static gravitational field reads 
\begin{equation}
L=-mc^{2}\sqrt{e^{2\phi }-\frac{V^{2}}{c^{2}}e^{-2\phi }}.  \label{u15}
\end{equation}%
Eq. (\ref{u12}) follows from Eq. (\ref{u11}) if multiply both sides of Eq. (%
\ref{u11}) by $\gamma e^{-2\phi }\mathbf{V}$ and make simple algebraic
transformations.

Lagrangian (\ref{u15}) gives the following expression for the particle
generalized momentum $\mathbf{p}=\frac{\partial L}{\partial \mathbf{V}}$%
\begin{equation}
\mathbf{p}=\gamma e^{-2\phi }m\mathbf{V},  \label{u16}
\end{equation}%
and particle Hamiltonian $H=\mathbf{V}\frac{\partial L}{\partial \mathbf{V}}%
\mathbf{-}L$%
\begin{equation}
H=e^{2\phi }\gamma mc^{2}=\sqrt{m^{2}c^{4}e^{2\phi }+p^{2}c^{2}e^{4\phi }}.
\label{u17}
\end{equation}%
Thus, Eq. (\ref{u12}) is the equation of energy conservation $E=$const,
where 
\begin{equation}
E=e^{2\phi }\gamma mc^{2}=\frac{e^{\phi }mc^{2}}{\sqrt{1-\frac{V^{2}}{c^{2}}%
e^{-4\phi }}}  \label{u19}
\end{equation}%
is the particle energy and Eq. (\ref{u11}) is the equation for momentum.

For a massless particle one should use Eq. (\ref{r14}) which for a static
field reads%
\begin{equation}
e^{-4\phi }\frac{\partial ^{2}\chi }{\partial t^{2}}-c^{2}\Delta \chi =0.
\label{u20}
\end{equation}%
Eq. (\ref{u20}) describes propagation of a massless particle with speed 
\begin{equation}
v=ce^{2\phi }.  \label{u21}
\end{equation}%
One can see that speed of light depends on the gravitational field $\phi $
and $v\leq c$ if $\phi $ is given by Eq. (\ref{r7}) with positive masses. By
proper rescaling of coordinates in Eq. (\ref{u20}) one can remove the factor 
$e^{-4\phi }$ at any given point. Let us fix $\phi =0$ at infinite distance
from masses. If an observer at infinity sends a light signal towards the Sun
then near the solar surface $\phi <0$ and light will propagate with a
smaller speed. This is the explanation of Shapiro time delay in the present
theory of gravity. In our theory the space-time geometry is fixed everywhere
and given by Minkowski metric. Light signal traveling the same distance
arrives with a delay if the light trajectory passes near the Sun. The delay
occurs because the speed of light is smaller near the solar surface.

Since Eq. (\ref{u20}) does not contain $t$ explicitly the photon frequency $%
\omega _{0}$ (measured in time $t$) remains the same during light
propagation. However, physical processes occur with different rates at
different $\phi $. Gravitational field (\ref{s10}) can be removed at a given
point by rescaling time in the factor $\sqrt{f_{00}}=e^{\phi }$ ($t=\tau
/e^{\phi }$) and spatial coordinates by $\sqrt{-f_{\alpha \alpha }}=e^{-\phi
}$. In such rescaled coordinates identical atoms emit light with equal
frequencies $\omega \propto $ $\partial \chi /\partial \tau =e^{-\phi
}\partial \chi /\partial t$. Thus we obtain 
\begin{equation}
\omega =\omega _{0}e^{-\phi }\text{,}  \label{p10}
\end{equation}%
where $\omega _{0}$ is the photon frequency measured in time $t$.

Eq. (\ref{p10}) shows that if light emitted by an atom propagates into a
region with larger gravitational potential then the detected light frequency
is smaller then those an identical atom emits at the detection point. This
phenomenon is known as gravitational redshift of light. Eq. (\ref{p10}) also
shows that in our theory there are no black holes. Indeed for the
gravitational field created by a point mass $M$: $\phi =-GM/c^{2}r$.
Therefore if a photon is emitted at a distance $r$ from the mass $M$ with
frequency $\omega $ then an observer at infinity will detect the photon with
the energy 
\begin{equation}
\hbar \omega _{0}=\hbar \omega e^{-GM/c^{2}r}\text{.}  \label{p11}
\end{equation}%
According to Eq. (\ref{p11}) no matter how close the photon is emitted to
the mass $M$ the photon's energy at infinity never becomes zero. This means
that photon can escape from the mass $M$ from any distance. Such a
conclusion is dramatically different from prediction of general relativity.
In Einstein's theory photons become trapped by the mass $M$ if they are
emitted from a distance smaller then the event horizon (that is point mass $%
M $ behaves as a black hole).

In Section \ref{test} we show that our theory passes all tests of general
relativity. In particular, Eqs. (\ref{u12}) and (\ref{u11}) explain
correctly the precession of the perihelion of Mercury and Eq. (\ref{p10})
the gravitational redshift of light.

\section{Stationary gravitational field}

Next we consider gravitational field produced by stationary mass currents,
so that the energy-momentum tensor of matter (\ref{s3}) is independent of
time. We assume that matter velocity $\mathbf{V}$ is much smaller then the
speed of light. However, the scalar potential $\phi $ is not necessarily
small. In this case one can look for solution for the equivalent metric (\ref%
{x3a}) in the form (in the Cartesian coordinate system)%
\begin{equation}
f_{ik}=\left( 
\begin{array}{cccc}
e^{2\phi } & \tilde{A}_{1}e^{2\phi } & \tilde{A}_{2}e^{2\phi } & \tilde{A}%
_{3}e^{2\phi } \\ 
\tilde{A}_{1}e^{2\phi } & -e^{-2\phi } & 0 & 0 \\ 
\tilde{A}_{2}e^{2\phi } & 0 & -e^{-2\phi } & 0 \\ 
\tilde{A}_{3}e^{2\phi } & 0 & 0 & -e^{-2\phi }%
\end{array}%
\right) ,  \label{z1}
\end{equation}%
where the three dimensional vector $\mathbf{\tilde{A}}=(\tilde{A}_{1},\tilde{%
A}_{2},\tilde{A}_{3})$ is small, $\tilde{A}\ll 1$, and $\phi =\phi (\mathbf{r%
})$ is arbitrary. For metric (\ref{z1}) the Ricci tensor upto terms linear
in $\mathbf{\tilde{A}}$ reads%
\begin{equation*}
R_{00}=e^{4\phi }\Delta \phi ,
\end{equation*}%
\begin{equation*}
R_{0\alpha }=e^{4\phi }\tilde{A}_{\alpha }\Delta \phi -\frac{1}{2}\left[ 
\text{curl}(e^{4\phi }\text{curl}\mathbf{\tilde{A})}\right] _{\alpha },
\end{equation*}%
\begin{equation*}
R_{\alpha \beta }=-2\frac{\partial \phi }{\partial x^{\alpha }}\frac{%
\partial \phi }{\partial x^{\beta }}+\Delta \phi \delta _{\alpha \beta }
\end{equation*}%
and Eqs. (\ref{e2}) reduce to

\begin{equation}
\Delta \phi =\frac{4\pi G}{c^{2}}e^{\phi }\rho ,\quad \nabla \phi \cdot 
\mathbf{\tilde{A}}=0,  \label{z3}
\end{equation}%
\begin{equation}
\text{curl}\left( e^{4\phi }\text{curl}\mathbf{\tilde{A}}\right) =\frac{%
16\pi G}{c^{3}}e^{\phi }\rho \mathbf{V,}  \label{z4}
\end{equation}%
here $\mathbf{V=}(V^{1},V^{2},V^{3})$ and $\rho $ are the matter velocity
and density respectively. Eqs. (\ref{z1}), (\ref{z3}) and (\ref{z4}) are
valid in the first order in $V/c$ and for $\mathbf{\tilde{A}=0}$ reduce to
those for a static field.

For a point mass $M$ rotating around the $z-$axis with an angular momentum $%
L $ Eqs. (\ref{z1}), (\ref{z3}) and (\ref{z4}) give the following expression
for $f_{ik}$ in spherical coordinates $(ct,r,\theta ,\varphi )$%
\begin{equation}
f_{ik}=\left( 
\begin{array}{cccc}
e^{2\phi } & 0 & 0 & Ar\sin \theta \\ 
0 & -e^{-2\phi } & 0 & 0 \\ 
0 & 0 & -r^{2}e^{-2\phi } & 0 \\ 
Ar\sin \theta & 0 & 0 & -r^{2}\sin ^{2}(\theta )e^{-2\phi }%
\end{array}%
\right) ,  \label{z9}
\end{equation}%
where%
\begin{equation}
\phi =-\frac{GM}{c^{2}r},\quad A=\frac{3c^{3}Lr\sin \theta }{16M^{3}G^{2}}%
\left[ e^{-2\phi }-e^{2\phi }\left( 8\phi ^{2}-4\phi +1\right) \right] .
\label{z10}
\end{equation}%
In the weak field limit $GM\ll c^{2}r$ Eq. (\ref{z10}) yields $A=2GL\sin
(\theta )/c^{3}r^{2}$ which coincides with the answer obtained in general
relativity.

One can find exact analytical solutions for gravitational field of any
strength by making Lorentz transformation of Eq. (\ref{s10}). For example,
gravitational field produced by a relativistic point particle moving with
constant velocity $\mathbf{V}$ along the $x-$axis is given by%
\begin{equation}
A_{k}=\frac{1}{\sqrt{c^{2}-V^{2}}}[c,-V,0,0],  \label{z5}
\end{equation}%
and 
\begin{equation}
\phi =-\frac{GM}{c^{2}\sqrt{\frac{(x-Vt)^{2}}{1-V^{2}/c^{2}}+y^{2}+z^{2}}},
\label{z6}
\end{equation}%
where $M$ is the particle rest mass.

\section{Tests of the theory of gravity}

\label{test}

According to the present theory we live in Minkowski space-time in which
there is vector gravitational field. However descriptions of particle motion
in terms of the vector field or space-time geometry are equivalent. This is
Einstein equivalence principle. One can consider interaction of particles
with the field in two different, but equivalent ways. In the first approach
the space-time is Minkowski space-time in which there is vector
gravitational field and motion of a test particle is described by Eq. (\ref%
{r10}). In the second treatment we describe interaction of the particle with
the gravitational field in a geometrical way. Namely we assume there is no
vector field, but instead the space-time is curved with metric%
\begin{equation}
g_{ik}=f_{ik},\quad g^{ik}=\tilde{f}^{ik}\text{,}  \label{k2}
\end{equation}%
and particles move along geodesic lines \cite{Land95}%
\begin{equation}
\frac{du^{b}}{ds}=-\Gamma _{ik}^{b}u^{i}u^{k},  \label{p12}
\end{equation}%
where $\Gamma _{ik}^{b}$ are Christoffel symbols, $ds=\sqrt{%
g_{ik}dx^{i}dx^{k}}$ and $u^{k}=dx^{k}/ds$ is the particle $4-$velocity.
Please note that Eq. (\ref{r10}) is written in Minkowski metric, while Eq. (%
\ref{p12}) is written in metric (\ref{k2}). However, mathematically
equations (\ref{r10}) and (\ref{p12}) are identical and thus give the same
answer.

It is convenient to compare the present theory of gravity with observations
based on the equivalent metric. According to Eq. (\ref{s10}) the equivalent
metric for static gravitational field is%
\begin{equation}
g_{ik}=\left( 
\begin{array}{cccc}
e^{2\phi } & 0 & 0 & 0 \\ 
0 & -e^{-2\phi } & 0 & 0 \\ 
0 & 0 & -e^{-2\phi } & 0 \\ 
0 & 0 & 0 & -e^{-2\phi }%
\end{array}%
\right) ,  \label{k6}
\end{equation}%
where for a point mass $M$: $\phi =-M/r$ (here we put $G=c=1$). Metric (\ref%
{k6}) is known as Yilmaz exponential metric \cite{Yilm58,Yilm71}. For a
point mass $M$ Einstein equations (\ref{i1}) yield Schwarzschild metric
which in isotropic Cartesian coordinates has the form \cite{Land95} 
\begin{equation}
ds^{2}=h(r)(dx^{0})^{2}-g(r)[(dx^{1})^{2}+(dx^{2})^{2}+(dx^{3})^{2}],
\label{p14}
\end{equation}%
where%
\begin{equation}
h(r)=\left( \frac{1-M/2r}{1+M/2r}\right) ^{2},\quad g(r)=\left( 1+\frac{M}{2r%
}\right) ^{4}.  \label{p15}
\end{equation}%
For small $M/r$ both the Schwarzschild (\ref{p15}) and Yilmaz (\ref{k6})
metrics yield the same expansion%
\begin{equation}
h(r)=1-\frac{2M}{r}+\frac{2M^{2}}{r^{2}}+\ldots ,\quad g(r)=1+\frac{2M}{r}%
+\ldots  \label{p16}
\end{equation}%
which is known as Post-Newtonian approximation. The four classic tests of
general relativity, namely the gravitational redshift of light, the
deflection of light by the Sun, the precession of the perihelion of Mercury
and time delay of a radar signal traveling near the Sun (Shapiro delay),
have examined the metric in the Post-Newtonian approximation (\ref{p16}) 
\cite{Will06}. Because the equivalent metric (\ref{k6}) obtained in the
present theory has correct Post-Newtonian limit (\ref{p16}) our theory of
gravity also passes the four classic tests. For static field the present
theory of gravity gives answer different from general relativity in the next
correction beyond the Post-Newtonian approximation. So far, however, gravity
have not been tested in this region.

In Appendix C we show that in our theory radiation of weak gravitational
waves is described by the same formula as in general relativity. Such
radiation was indirectly detected as energy loss by binary pulsars and
served as a quantitative test of Einstein equations for weak time-dependent
field. The present theory also passes this test.

\section{Cosmology}

\label{cosmo}

In this section we apply our theory to evolution of the Universe. We assume
that one can omit kinetic energy of matter compared to its rest energy. We
also assume that matter is uniformly distributed in space with density $\rho 
$, where $\rho $ is independent of time. According to the present theory we
live in Minkowski space-time and galaxies located in different parts of the
Universe do not move relative to each other (apart from local random motion
which we omit in this section). That's why $\rho $ is constant. However
gravitational field evolves with time. Light emitted by a distant source
reaches an observer on Earth with a delay. As a result, at the moment of
light detection the gravitational field is different from its value when
light was emitted. This is the origin of cosmological redshift.

For spatially isotropic Universe gravitational field must have the form 
\begin{equation}
A_{k}=[1,0,0,0]  \label{c1}
\end{equation}%
and, thus, equivalent metric (\ref{x3a}) is given by 
\begin{equation}
f_{ik}=\left( 
\begin{array}{cccc}
\frac{1}{a^{2}} & 0 & 0 & 0 \\ 
0 & -a^{2} & 0 & 0 \\ 
0 & 0 & -a^{2} & 0 \\ 
0 & 0 & 0 & -a^{2}%
\end{array}%
\right) ,  \label{c2}
\end{equation}%
where $a=e^{-\phi }$. As in general relativity, we add the cosmological
constant term to energy-momentum tensor, $T_{ik}\rightarrow T_{ik}+\Lambda
f_{ik}$, where $\Lambda $ is a constant. Then for gravitational field (\ref%
{c1}) Eqs. (\ref{e2}) read%
\begin{equation}
R_{i0}=\frac{8\pi G}{c^{4}}\left( T_{i0}-\frac{1}{2}f_{i0}T-\Lambda
f_{i0}\right) .  \label{c3}
\end{equation}%
We also assume that the Universe is homogeneous on large spatial scales and,
therefore, $a$ in Eq. (\ref{c2}) must be independent of $\mathbf{r}$. Then
for the equivalent metric (\ref{c2}) the Ricci tensor is 
\begin{equation}
R_{00}=-\frac{3}{a^{2}}\frac{\partial }{\partial t}\left( a\dot{a}\right)
,\quad R_{\alpha 0}=0,  \label{c4}
\end{equation}%
and%
\begin{equation}
R_{\alpha \beta }=\delta _{\alpha \beta }\frac{\partial }{\partial t}\left(
a^{3}\dot{a}\right) .  \label{c5}
\end{equation}%
The energy-momentum tensor of matter (\ref{s3}) has only one nonzero
component $T_{00}=\rho c^{2}/a^{5}$, $T=\rho c^{2}/a^{3}$, and therefore
Eqs. (\ref{c3}) reduce to 
\begin{equation}
-3\frac{\partial }{\partial t}\left( a\dot{a}\right) =\frac{4\pi G}{c^{4}}%
\left( \frac{\rho c^{2}}{a^{3}}-2\Lambda \right) .  \label{c6}
\end{equation}%
Multiplying both sides of Eq. (\ref{c6}) by $a\dot{a}$ and integrating over
time we obtain%
\begin{equation}
3\dot{a}^{2}=\frac{8\pi G}{c^{4}}\left( \frac{\rho c^{2}}{a^{3}}+\Lambda +%
\frac{C}{a^{2}}\right) ,  \label{w5}
\end{equation}%
where $C$ is an integration constant. For $\Lambda =0$ sign of $C$
determines whether a Universe is open or closed. In our theory the constant $%
C$ produces the same effect on $a(t)$ as the curvature of space in general
relativity.

In the metric (\ref{c2}) Einstein equations (\ref{i1}) with the cosmological
constant term give our Eq. (\ref{w5}) with $C=0$. Indeed, present Eqs. (\ref%
{c3}) coincide with Einstein equations (\ref{i1}) with index $k=0$. In
addition, Einstein equations with $ik=\alpha \alpha $ yield 
\begin{equation}
\frac{\partial }{\partial t}\left( a^{3}\dot{a}\right) =\frac{8\pi G}{c^{4}}%
\left( \frac{\rho c^{2}}{2a}+\Lambda a^{2}\right) .  \label{c7}
\end{equation}%
Multiplying both sides of Eq. (\ref{c7}) by $a^{3}\dot{a}$ and integrating
over time we obtain%
\begin{equation}
3\dot{a}^{2}=\frac{8\pi G}{c^{4}}\left( \frac{\rho c^{2}}{a^{3}}+\Lambda +%
\frac{C_{1}}{a^{6}}\right) ,  \label{c8}
\end{equation}%
where $C_{1}$ is an integration constant. Eqs. (\ref{w5}) and (\ref{c8}) are
compatible if $C_{1}=C=0$, and thus our Eq. (\ref{w5}) with $C=0$ is also
solution in general relativity. Such a solution agrees with available
cosmological data.

By changing time coordinate into 
\begin{equation}
\tau =\int^{t}\frac{dt}{a(t)}  \label{y9}
\end{equation}%
Eq. (\ref{w5}) yields%
\begin{equation}
\frac{3}{a^{2}}\left( \frac{\partial a}{\partial \tau }\right) ^{2}=\frac{%
8\pi G}{c^{4}}\left( \frac{\rho c^{2}}{a^{3}}+\Lambda +\frac{C}{a^{2}}\right)
\label{y10}
\end{equation}%
which describes evolution of the Universe in the metric $ds^{2}=c^{2}d\tau
^{2}-a^{2}(dx^{2}+dy^{2}+dz^{2})$. Such form of metric is commonly used in
cosmology for spatially flat Universe.

\section{Galactic centers and dark matter problem}

\label{dark}

In the present theory static gravitational field is described by the
equivalent exponential metric (\ref{k6}). Metric (\ref{k6}) was also
obtained in Refs. \cite{Yilm58,Yilm71,Rose71,Chan80}. Exponential metric (%
\ref{k6}) predicts no black holes, but rather compact objects with no event
horizon and very large, but finite, gravitational redshift.

In recent years, the evidence for the existence of an ultra-compact
concentration of dark mass at centers of galaxies has become very strong.
However, a proof that such objects are black holes rather then compact
objects without event horizon is lacking. If the present theory of gravity
is correct then the compact supermassive objects at galactic centers can not
be composed of baryonic matter. Indeed, mass of a compact (neutron star
like) baryonic object in the exponential metric (\ref{k6}) can not exceed
about $12$M$_{\odot }$ \cite{Robe99}, but the objects at galactic centers
possess masses upto a few $10^{9}$M$_{\odot }$. Hence, those objects must be
made of dark matter of non baryonic origin. This fact gives us an
opportunity to determine composition of dark matter based on observations of
supermassive objects at galactic centers.

In the previous paper \cite{Svid07} we showed that properties of compact
objects at galactic centers can be explained quantitatively assuming they
are made of dark matter axions and the axion mass is about $0.6$ meV.
Analysis of Ref. \cite{Svid07} is based on the assumption that static
gravitational field is described by the exponential metric (\ref{k6}) rather
then by general relativity. A full time-dependent theory of gravity was
unnecessary for calculations made in Ref. \cite{Svid07}. The present paper
provides such a theory and justifies our previous choice of the exponential
metric.

Axions are one of the leading particle candidates for the cold dark matter
in the Universe \cite{Brad03}. Interaction of axions with QCD instantons
generates the axion mass $m$ and periodic interaction potential \cite{Kim87} 
\begin{equation}
V(\varphi )=m^{2}F^{2}[1-\cos (\varphi /F)],  \label{p1}
\end{equation}%
where $\varphi $ is a real scalar axion field and $F$ is the Peccei-Quinn
symmetry breaking scale. The interaction potential (\ref{p1}) has degenerate
minima $V=0$ at $\varphi =2\pi nF$, where $n$ is an integer number. As a
consequence, axions can form bubbles. Bubble mass is concentrated in a thin
surface (interface between two degenerate vacuum states). In the exponential
metric the potential energy of a spherical bubble with radius $R$ is given
by \cite{Svid07}%
\begin{equation}
U(R)=4\pi \sigma R^{2}\exp \left( \frac{M}{R}\right) ,
\end{equation}%
where $\sigma $ is the surface energy density and $M$ is the fixed bubble
mass. $U(R)$ has a shape of a well. At $R\gg M$ one can omit gravity and $%
U(R)\simeq 4\pi \sigma R^{2}$ is just a surface energy (tension) which tends
to contract the bubble. At $R\ll M$ gravity effectively produces large
repulsive potential which forces the bubble to expand. As a result, the
bubble radius $R(t)$ oscillates between two turning points.

In Ref. \cite{Svid07}, based on quantitative analysis of available data, we
argued that such oscillating axion bubbles, rather then supermassive black
holes, could be present at galactic centers. Recent observations of
near-infrared and X-ray flares from Sagittarius A$^{\ast }$, which is
believed to be a $3.6\times 10^{6}$M$_{\odot }$ black hole at the Galactic
center, show that the source exhibits about $20$-minute periodic variability 
\cite{Genz03,Gill06,Bela06}. An oscillating axion bubble can explain such
variability. Known value of the bubble mass at the center of our Galaxy and
its oscillation period yields the axion mass of about $0.6$ meV. Size of the
axion bubble at the center of the Milky Way oscillates between $R_{\min
}\approx 1R_{\odot }$ and $R_{\max }\approx 1$AU $\approx 210R_{\odot }$.

Further, as shown in Ref. \cite{Svid07}, the axion bubbles with no free
parameters (if we fix $m=0.6$ meV based on Sagittarius A* flare variability)
quantitatively explain the upper limit (a few $10^{9}$M$_{\odot }$) on the
supermassive \textquotedblleft black hole" mass found in recent analysis of
the measured mass distribution \cite{Grah07}. Also, with no free parameters
the bubble scenario explains observed lack of supermassive \textquotedblleft
black holes" with mass $M\lesssim 10^{6}$M$_{\odot }$ \cite{Grah08}. For
such low-mass bubbles the decay time $t\propto M^{9/2}$ becomes much shorter
then the age of the Universe and, as a result, such objects are very rare.

Observation of the Galactic center with very long-baseline interferometry
within the next few years will be capable to test theories of gravitation in
the strong field limit. Such an observation will allow us to distinguish
between the black hole (predicted by general relativity) and the oscillating
axion bubble scenario. A defining characteristic of a black hole is the
event horizon. To a distant observer, the event horizon casts a relatively
large \textquotedblleft shadow\textquotedblright\ over the background source
with an apparent diameter of about $10GM/c^{2}\approx 80R_{\odot }$ due to
bending of light. The predicted size of this shadow for Sagittarius A*
approaches the resolution of current radio-interferometers. Hence, there
exists a realistic expectation of imaging the shadow of a black hole with
VLBA within the next few years \cite{Falc00,Falc01,Shen05,Huan07,Doel09}. If
the axion bubble, rather then a black hole, is present at the Galactic
center, the steady shadow will not be observed. Instead, the shadow will
appear and disappear periodically with a period of about $20$ $\min $.
Discovery of periodic appearance of the shadow from the Galactic center
object will also be a strong evidence for the axion nature of dark matter
and will lead to an accurate prediction of the axion mass.

One should mention that intrinsic size of Sagittarius A* at a wavelength of $%
1.3$ mm was recently determined using VLBA \cite{Doel08}. The intrinsic
diameter of Sagittarius A* was found to be $<0.3$AU$\approx 65R_{\odot }$
which is less than the expected apparent size of the event horizon of the
presumed black hole. Such observation might indicate lack of black holes, in
agreement with the present theory.

\section{Conclusions}

Here we propose a new classical theory of gravity which is based on the
principle of equivalence and assumption that, similarly to electrodynamics,
gravity is described by a vector field in Minkowski space-time. We show that
present theory is the only possibility that can be obtained from these
assumptions. Our theory fundamentally differs from general relativity which
treats space-time geometry as gravitational field. In the present theory,
similarly to the Standard Model, matter does not affect geometry of flat
Minkowski space-time.

The current vector theory is not equivalent to general relativity even in
the weak field limit. Nevertheless, our theory also passes all available
tests and for static field in the Post-Newtonian approximation gives the
same answer as general relativity. Beyond the Post-Newtonian approximation
the present theory gives different result and yields no singularities such
as black holes. A defining characteristic of a black hole is the event
horizon. So far there were no observations of the event horizon and, thus, a
proof of black holes existence is lacking. For cosmology our theory predicts
essentially the same evolution of the Universe as general relativity.

In the present theory gravitational field is described by four equations (%
\ref{e2}) which can be solved analytically for much greater number of
problems then ten Einstein equations (\ref{i1}). In particular, for
arbitrary static mass distribution our Eqs. (\ref{e2}) have exact analytical
solution (\ref{r7}).

The present theory, if confirmed, can also lead to a break through in the
problem of dark matter. Namely, the theory predicts that supermassive
compact objects at galactic centers have non baryonic origin and, thus, yet
undiscovered dark matter particle is a likely ingredient for their
composition. As a result, observations of such objects can allow us to
predict the nature of dark matter. In the previous paper \cite{Svid07} we
showed that properties of compact objects at galactic centers can be
explained quantitatively assuming they are made of dark matter axions and
the axion mass is about $0.6$ meV. Analysis of Ref. \cite{Svid07} is based
on the present exponential metric (\ref{k6}) for static gravitational field
rather then general relativity.

Our theory of gravity can be tested in several ways. For example, one can
examine gravity beyond the Post-Newtonian approximation in the solar system
by improving the accuracy of Shapiro time delay experiment (time delay of a
radar signal traveling near the Sun). Another possibility is to resolve the
supermassive object at the center of our Galaxy with VLBA. If general
relativity is correct we must see a steady shadow from a black hole. If the
present theory is right then likely the shadow will appear and disappear
periodically with a period of about $20$ min as we predicted in \cite{Svid07}%
. Observation of such oscillations will also provide evidence for dark
matter axion with mass in meV range.

This work was supported by the Office of Naval Research (Award No.
N00014-07-1-1084 and N0001408-1-0948).

\appendix

\section{Derivation of equivalent metric}

Here we obtain expression for the metric $f_{ik}$ which is equivalent to the
vector field $A_{k}$ using the principle of equivalence. For small $A_{k}$
one can expand $f_{ik}$ in powers of $A_{k}$. In the leading order there are
two possible terms $A_{i}A_{k}$ and $A^{2}\eta _{ik}$, where $\eta _{ik}=$%
diag$(1,-1,-1,-1)$ is Minkowski metric tensor and $A^{2}=A_{k}A^{k}$ is
square of vector $A_{k}$ ($A^{i}=\eta ^{ik}A_{k}$). To obey available
experimental tests these terms must enter $f_{ik}$ in the combination $%
A^{2}\eta _{ik}-2A_{i}A_{k}$ and, therefore, $f_{ik}$ has the form 
\begin{equation}
f_{ik}\approx \eta _{ik}+A^{2}\eta _{ik}-2A_{i}A_{k}\text{.}  \label{u4}
\end{equation}%
One can write Eq. (\ref{u4}) as 
\begin{equation}
f_{ik}\approx \eta _{ik}+\eta _{im}\phi _{k}^{m}\text{,}  \label{u4a}
\end{equation}%
where 
\begin{equation}
\phi _{k}^{m}=A^{2}\delta _{k}^{m}-2A^{m}A_{k}\text{.}  \label{u10}
\end{equation}%
In curved space-time with metric $g_{ik}$ Eq. (\ref{u4a}) reads%
\begin{equation}
f_{ik}\approx g_{ik}+g_{im}\phi _{k}^{m},  \label{q5}
\end{equation}%
where $\phi _{k}^{m}$ is written in metric $g_{ik}$.

Let us now assume that gravitational field is not small. According to the
equivalence principle one can describe motion of particles as if there is no
field, but instead the space-time is curved with an equivalent metric $%
g_{ik} $. If we change $\phi _{ik}$ by a small amount $\phi _{ik}\rightarrow
\phi _{ik}+\alpha \phi _{ik}$, where $\alpha $ is a small number, then we
can write ($\phi \equiv \phi _{ik}$) 
\begin{equation}
f_{ik}(\phi +\alpha \phi )\approx f_{ik}(\phi )+\left. \frac{\partial
f_{ik}[(1+\alpha )\phi ]}{\partial \alpha }\right\vert _{\alpha =0}\alpha .
\label{u1}
\end{equation}%
From the other hand Eq. (\ref{q5}) yields 
\begin{equation}
f_{ik}(\phi +\alpha \phi )\approx f_{ik}(\phi )+f_{im}(\phi )\alpha \phi
_{k}^{m}.  \label{u2}
\end{equation}%
Comparing Eqs. (\ref{u1}) and (\ref{u2}) we obtain that for any $\phi _{ik}$ 
\begin{equation}
\left. \frac{\partial f_{ik}[(1+\alpha )\phi ]}{\partial \alpha }\right\vert
_{\alpha =0}=f_{im}(\phi )\phi _{k}^{m}.  \label{u3}
\end{equation}%
Because Eq. (\ref{u3}) must be satisfied for any $\phi _{ik}$ it is
essentially a differential equation, rather then a condition at one point ($%
\alpha =0$). Solution of Eq. (\ref{u3}) yields the following unique
expression for the metric equivalent to the gravitational field $\phi _{ik}$%
\begin{equation}
f_{ik}=\eta _{im}\exp (\phi _{k}^{m}),  \label{u5}
\end{equation}%
where the exponent of a tensor is defined as%
\begin{equation}
\exp (\phi _{k}^{m})=\delta _{k}^{m}+\phi _{k}^{m}+\frac{1}{2!}\phi
_{l}^{m}\phi _{k}^{l}+\frac{1}{3!}\phi _{l}^{m}\phi _{p}^{l}\phi
_{k}^{p}+\ldots .  \label{t5}
\end{equation}%
Substituting Eq. (\ref{u10}) into Eq. (\ref{u5}) and taking into account
that 
\begin{equation}
\exp (-2A^{l}A_{k})=\delta _{k}^{l}-\frac{A^{l}A_{k}}{A^{2}}\left(
1-e^{-2A^{2}}\right)
\end{equation}%
we obtain%
\begin{equation*}
f_{ik}=\eta _{im}\exp (A^{2}\delta _{k}^{m}-2A^{m}A_{k})=e^{A^{2}}\eta
_{im}\exp (-2A^{m}A_{k})
\end{equation*}%
\begin{equation}
=e^{A^{2}}\eta _{ik}-\frac{2\eta _{im}A^{m}A_{k}}{A^{2}}\sinh (A^{2}).
\label{t6}
\end{equation}%
In Eq. (\ref{t6}) $A_{k}$ is the gravitational field in metric $f_{ik}$.
Therefore, $A_{i}=f_{ik}A^{k}$. Taking into account Eq. (\ref{t6}) we find%
\begin{equation}
A_{i}=e^{-A^{2}}\eta _{ik}A^{k},\quad A^{i}=e^{A^{2}}\eta ^{ik}A_{k},
\label{t7}
\end{equation}%
and $A^{2}=A_{k}A^{k}$ is determined from the equation%
\begin{equation}
A^{2}e^{-A^{2}}=\eta ^{ik}A_{i}A_{k}.  \label{t8}
\end{equation}%
As a result one can write $f_{ik}$ as 
\begin{equation}
f_{ik}=e^{A^{2}}\eta _{ik}-\frac{A_{i}A_{k}}{A^{2}}(e^{2A^{2}}-1).
\end{equation}%
It is convenient to rewrite this equation in a different form introducing $%
\tilde{A}_{k}=A_{k}e^{A^{2}/2}$. Then 
\begin{equation}
f_{ik}=e^{\tilde{A}^{2}}\eta _{ik}-2\frac{\tilde{A}_{i}\tilde{A}_{k}}{\tilde{%
A}^{2}}\sinh (\tilde{A}^{2}),
\end{equation}%
where 
\begin{equation}
\tilde{A}^{2}=\eta ^{ik}\tilde{A}_{i}\tilde{A}_{k}.
\end{equation}

\section{Motion of particles in gravitational field}

\label{motion}

Here we find how a test particle with rest mass $m$ moves in an external
gravitational field $f_{ik}$. Interaction of the particle with the field is
described by the action 
\begin{equation}
I_{\text{field-matter}}=-mc\int \sqrt{f_{ik}dx^{i}dx^{k}},  \label{r8}
\end{equation}%
where the integral is taken along the particle trajectory. One can find
equation of particle motion varying the action (\ref{r8}) at fixed $f_{ik}$ 
\cite{Land95}%
\begin{equation}
\delta I_{\text{field-matter}}=-\frac{mc}{2}\int \left[ u^{i}u^{k}ds\delta
f_{ik}+f_{ik}(u^{k}d\delta x^{i}+u^{i}d\delta x^{k})\right] ,
\end{equation}%
where $u^{i}$ is the particle 4-velocity%
\begin{equation}
u^{i}=\frac{dx^{i}}{ds}  \label{r11}
\end{equation}%
with $ds=\sqrt{f_{ik}dx^{i}dx^{k}}$. Next we take into account $\delta
f_{ik}=\left( \partial f_{ik}/\partial x^{l}\right) \delta x^{l}$ and
integrate the second term by parts%
\begin{equation*}
\delta I_{\text{field-matter}}=
\end{equation*}%
\begin{equation*}
-\frac{mc}{2}\int \left[ \left( \frac{\partial f_{ik}}{\partial x^{l}}-\frac{%
\partial f_{lk}}{\partial x^{i}}-\frac{\partial f_{il}}{\partial x^{k}}%
\right) u^{i}u^{k}-2f_{lk}\frac{du^{k}}{ds}\right] ds\delta x^{l}.
\end{equation*}%
Principle of least action $\delta I_{\text{field-matter}}=0$ yields the
following equation%
\begin{equation}
f_{lk}\frac{du^{k}}{ds}=\frac{1}{2}\left[ \frac{\partial f_{ik}}{\partial
x^{l}}-\frac{\partial f_{lk}}{\partial x^{i}}-\frac{\partial f_{il}}{%
\partial x^{k}}\right] u^{i}u^{k}.  \label{r9}
\end{equation}%
Multiplying both sides of Eq. (\ref{r9}) by tensor inverse to $f_{lk}$ we
find%
\begin{equation}
\frac{du^{b}}{ds}=\frac{1}{2}\tilde{f}^{bl}\left[ \frac{\partial f_{ik}}{%
\partial x^{l}}-\frac{\partial f_{lk}}{\partial x^{i}}-\frac{\partial f_{il}%
}{\partial x^{k}}\right] u^{i}u^{k},  \label{r10}
\end{equation}%
or 
\begin{equation}
\frac{du^{b}}{ds}=-\Gamma _{ik}^{b}u^{i}u^{k}.  \label{s9}
\end{equation}%
This is equation of motion of a particle in gravitational field $f_{ik}$.

From Eq. (\ref{r8}) we obtain the following Lagrangian of the particle 
\begin{equation}
L=-mc\sqrt{f_{ik}\frac{dx^{i}}{dt}\frac{dx^{k}}{dt}}.  \label{r12}
\end{equation}

Action (\ref{r8}) and Eq. (\ref{r10}) are invalid for massless particles.
Let us consider a massless scalar field $\chi $. In the gravitational field $%
f_{ik}$ the action for $\chi $ reads (in Minkowski metric)%
\begin{equation}
I=\frac{1}{8\pi }\int d^{4}x\sqrt{-f}\tilde{f}^{\mu \nu }\frac{\partial \chi
^{\ast }}{\partial x^{\mu }}\frac{\partial \chi }{\partial x^{\nu }}.
\label{r13}
\end{equation}%
Variation of Eq. (\ref{r13}) yields the following equation of motion for the
field $\chi $%
\begin{equation}
\frac{\partial }{\partial x^{\mu }}\left( \sqrt{-f}\tilde{f}^{\mu \nu }\frac{%
\partial \chi }{\partial x^{\nu }}\right) =0.  \label{r14}
\end{equation}%
For geometrical optics one can write $\chi $ as $\chi =|\chi |e^{i\psi }$,
where $\psi $ (eikonal) has a large value. Substituting this into Eq. (\ref%
{r14}) and keeping only the leading term we obtain eikonal equation in
gravitational field 
\begin{equation}
\tilde{f}^{\mu \nu }\frac{\partial \psi }{\partial x^{\mu }}\frac{\partial
\psi }{\partial x^{\nu }}=0.  \label{r15}
\end{equation}

\section{Radiation of gravitational waves}

Here we consider radiation of weak gravitational waves. In general
relativity the metric for weak plane gravitational waves propagating along
the $x-$axis in a properly chosen coordinate system reads \cite{Land95}%
\begin{equation}
g_{ik}=\eta _{ik}+\left( 
\begin{array}{cccc}
0 & 0 & 0 & 0 \\ 
0 & 0 & 0 & 0 \\ 
0 & 0 & h_{22}(t,x) & h_{23}(t,x) \\ 
0 & 0 & h_{23}(t,x) & -h_{22}(t,x)%
\end{array}%
\right) ,  \label{wa3}
\end{equation}%
where $h_{23}$ and $h_{22}$ are small perturbations. In linear approximation
the corresponding Ricci tensor is%
\begin{equation}
R_{ik}=\frac{1}{2}\left( 
\begin{array}{cccc}
0 & 0 & 0 & 0 \\ 
0 & 0 & 0 & 0 \\ 
0 & 0 & \square h_{22} & \square h_{23} \\ 
0 & 0 & \square h_{23} & -\square h_{22}%
\end{array}%
\right) .  \label{wa4}
\end{equation}%
To find energy radiated in the $x-$direction one should find functions $%
h_{23}$ and $h_{22}$ in the wave zone. They are determined from Einstein
equations (\ref{i1})%
\begin{equation}
\frac{1}{2}\square h_{22}=\frac{8\pi G}{c^{4}}\left( T_{22}-\frac{1}{2}%
g_{22}T\right) ,  \label{w6}
\end{equation}%
\begin{equation}
\frac{1}{2}\square h_{23}=\frac{8\pi G}{c^{4}}T_{23},  \label{w7}
\end{equation}%
\begin{equation}
-\frac{1}{2}\square h_{22}=\frac{8\pi G}{c^{4}}\left( T_{33}-\frac{1}{2}%
g_{33}T\right) .  \label{w8}
\end{equation}%
In the right hand side of Eqs. (\ref{w6})-(\ref{w8}) one can replace $g_{ik}$
with Minkowski metric $\eta _{ik}$.

In the present theory of gravity for weak gravitational waves ($|\phi |\ll 1$%
) propagating along the $x-$axis%
\begin{equation}
A_{k}=i[0,0,\sin \alpha ,\cos \alpha ]  \label{wa1}
\end{equation}%
and the equivalent metric (\ref{x3a}) has the form 
\begin{equation}
f_{ik}=\eta _{ik}+2\phi \left( 
\begin{array}{cccc}
-1 & 0 & 0 & 0 \\ 
0 & 1 & 0 & 0 \\ 
0 & 0 & \cos (2\alpha ) & -\sin (2\alpha ) \\ 
0 & 0 & -\sin (2\alpha ) & -\cos (2\alpha )%
\end{array}%
\right) ,  \label{wa2}
\end{equation}%
where $\phi =\phi (x-ct)$ and $\alpha =\alpha (x-ct)$. Eqs. (\ref{wa1}) and (%
\ref{wa2}) are equations in Minkowski space time. Let us make the following
change of coordinates%
\begin{equation}
t\rightarrow t+\frac{1}{c}\int_{0}^{x-ct}\phi (q)dq,  \label{w11}
\end{equation}%
\begin{equation}
x\rightarrow x-\int_{0}^{x-ct}\phi (q)dq.  \label{w12}
\end{equation}%
Taking into account that under coordinate transformation $x^{i}\rightarrow
x^{i}+\xi ^{i}$, where $\xi ^{i}$ are small, tensor $f_{ik}$ transforms as $%
f_{ik}\rightarrow f_{ik}-\partial \xi _{i}/\partial x^{k}-\partial \xi
_{k}/\partial x^{i}$ \cite{Land95} we obtain that in new coordinates 
\begin{equation}
f_{ik}=\eta _{ik}+2\phi \left( 
\begin{array}{cccc}
0 & 0 & 0 & 0 \\ 
0 & 0 & 0 & 0 \\ 
0 & 0 & \cos (2\alpha ) & -\sin (2\alpha ) \\ 
0 & 0 & -\sin (2\alpha ) & -\cos (2\alpha )%
\end{array}%
\right)  \label{w14}
\end{equation}%
which is precisely Eq. (\ref{wa3}) with $h_{22}=2\phi \cos (2\alpha )$ and $%
h_{23}=-2\phi \sin (2\alpha )$. As a result, for weak gravitational waves
our equivalent metric $f_{ik}$ in the wave zone is equal to $g_{ik}$, where $%
g_{ik}$ is solution of Einstein equations (\ref{w6})-(\ref{w8}). Indeed, if $%
f_{ik}$ obeys Eqs. (\ref{w6})-(\ref{w8}) then it automatically satisfies
present equations 
\begin{equation}
R_{ik}A^{k}=\frac{8\pi G}{c^{4}}\left( T_{ik}-\frac{1}{2}f_{ik}T\right) A^{k}
\label{w13}
\end{equation}%
which are linear combinations of Eqs. (\ref{w6})-(\ref{w8}) with
coefficients $A^{k}$.

We obtain that general relativity and our theory yield the same answer for
the equivalent metric of emitted weak gravitational waves. Therefore, in
both theories the energy loss due to emission of weak gravitational waves is
given by the same formula.

\section{Weak gravitational field}

Here we obtain equations for weak gravitational field. In this case $|\phi
|\ll 1$, however, the unit vector $A_{k}$ can be arbitrary. For weak field
the Ricci tensor (\ref{Rt}) reduces to 
\begin{equation}
R_{ik}\approx \frac{\partial \Gamma _{ik}^{l}}{\partial x^{l}}-\frac{%
\partial \Gamma _{il}^{l}}{\partial x^{k}}.
\end{equation}%
Taking into account Eq. (\ref{Cs}) for $\Gamma _{ik}^{l}$ and formula $%
\Gamma _{il}^{l}=\partial \ln \sqrt{-f}/\partial x^{i}=-2\partial \phi
/\partial x^{i}$ \cite{Land95} we find that Eqs. (\ref{e2}) in the weak
field limit yield%
\begin{equation*}
2A^{k}\left[ \frac{\partial ^{2}}{\partial x^{l}\partial x^{k}}\left(
A_{i}A^{l}\phi \right) +\frac{\partial ^{2}}{\partial x^{l}\partial x^{i}}%
\left( A_{k}A^{l}\phi \right) +\square (A_{i}A_{k}\phi )\right] -
\end{equation*}%
\begin{equation}
-A_{i}\square \phi =\frac{8\pi G}{c^{4}}\left( T_{ik}-\frac{1}{2}\eta
_{ik}T\right) A^{k},  \label{e2w}
\end{equation}%
where 
\begin{equation*}
\square \equiv \Delta -\frac{1}{c^{2}}\frac{\partial ^{2}}{\partial t^{2}},
\end{equation*}%
$T_{ik}$ is the energy momentum tensor for free particles 
\begin{equation}
T_{ik}=\sqrt{1-\frac{V^{2}}{c^{2}}}\rho c^{2}u_{i}u_{k},\quad T=\sqrt{1-%
\frac{V^{2}}{c^{2}}}\rho c^{2},
\end{equation}%
$\rho $ is the rest mass density and $u_{k}$ is the particle $4-$velocity 
\begin{equation}
u_{k}=\frac{1}{\sqrt{1-\frac{V^{2}}{c^{2}}}}\left( 1,\frac{V_{\alpha }}{c}%
\right) .
\end{equation}%
The weak field equations (\ref{e2w}) are Lorentz covariant and nonlinear in $%
A_{k}$. Eqs. (\ref{e2w}) differ from the weak field limit of Einstein
equations \cite{Land95}. This is the case because the present vector theory
of gravity is not equivalent to general relativity (tensor theory) even for
weak field.

If $A_{\alpha }$ are also small ($|A_{\alpha }|\ll 1$ and $A_{0}\approx 1$)
one can rewrite Eqs. (\ref{e2w}) in a form similar to Maxwell's equations
(for consistency we assume that $V\ll c$ and omit second order time
derivatives) 
\begin{equation}
\text{div}\mathbf{E}=4\pi \frac{G\rho }{c^{2}},
\end{equation}%
\begin{equation}
\text{curl}\mathbf{B}=\frac{4\pi }{c}\frac{G\rho \mathbf{V}}{c^{2}}+\frac{1}{%
c}\frac{\partial \mathbf{E}}{\partial t},
\end{equation}%
where 
\begin{equation*}
\mathbf{E}=\nabla \phi ,\quad \mathbf{B=}\text{curl}\mathbf{(A}\phi ),
\end{equation*}%
$\mathbf{A=}(A^{1},A^{2},A^{3})$ and $\mathbf{V=}(V^{1},V^{2},V^{3})$ is the
three dimensional velocity of particle.

\end{document}